\documentstyle[12pt]{article} 
\title{\baselineskip=9mm
Importance of Non-Linear Couplings in Fusion Barrier Distributions 
and Mean Angular Momenta \\}
\author{K. Hagino$^{1}$, N. Takigawa$^{1}$,
M. Dasgupta$^{2}$, D.J. Hinde$^{2}$, and J.R. Leigh$^{2}$
\\ \\
\medskip
{\it $^{1}$ Department of Physics,
Tohoku University, Sendai 980--77, Japan}
\\
{\it $^{2}$ Department of Nuclear Physics,} \\
{\it Research School of Physical
Sciences and Engineering, } \\
 {\it Australian National University, Canberra,
ACT 0200, Australia}
}
\date{}

\begin{document}

\maketitle

\vspace*{-1cm}

\begin{center}
{\bf Abstract}
\end{center}

The effects of higher order coupling of surface vibrations to 
the relative motion on heavy-ion fusion reactions 
at near-barrier energies are investigated. 
The coupled channels 
equations are solved to all orders, and also in the 
linear and the quadratic coupling approximations. 
It is shown 
that the shape of fusion barrier distributions and the energy dependence 
of the average angular momentum of the compound nucleus 
can significantly change when the 
higher order couplings are included. 
The role of octupole vibrational excitation of $^{16}$O 
in the $^{16}$O + $^{144}$Sm fusion reaction is also discussed 
using the all order coupled-channels equations. 

\medskip

\begin{flushleft} 
{\bf 1. Introduction}
\end{flushleft}

The analysis of the fusion process in terms of the barrier distribution 
has generated renewed interest 
in heavy-ion collisions at energies below and near the Coulomb 
barrier\cite{LDH95}. 
In a simple eigenchannel approach, couplings of the relative motion of the 
colliding nuclei to nuclear intrinsic degrees of freedom result in 
the single fusion barrier being replaced by a distribution 
of potential barriers. A method of extracting barrier distributions 
directly from fusion excitation functions was proposed\cite{RSS91}, and 
stimulated precise measurements of the fusion cross sections for 
several systems. 
The analysis of the barrier distribution clearly shows 
the effects of couplings to 
several nuclear intrinsic motions 
in a way much more apparent than in the fusion excitation function 
itself\cite{LDH95}. 

Theoretically the standard way to address the effects of the 
coupling between 
the relative motion and the intrinsic degrees of freedom is to 
solve the coupled-channels equations, including all the relevant channels. 
Most of the coupled channels calculations performed so far use 
the linear coupling approximation, where the 
coupling potential is expanded in powers of the deformation 
parameter, keeping only the linear term. 
Whilst this approach reproduces the experimental 
data of fusion cross sections for very asymmetric systems, 
it does not explain the data for heavier and nearly symmetric 
systems\cite{CBD95}. 
Thus, it is of interest to examine 
the validity of one of the main approximations in these calculations, namely 
the linear coupling approximation, and see whether the effects 
of non-linear coupling improve the agreement between data and 
the theoretical calculations for such systems. 
Even in asymmetric systems, the non-linear couplings might be 
important to reproduce precisely measured data. 

In this contribution, we report the results of the coupled channels 
calculations which include the couplings to all orders\cite{HTDHL97,BBK94}. 
The results of these calculations for fusion cross sections, average angular 
momenta and barrier distributions are compared with 
those using the linear and the quadratic coupling approximations. 
It is seen that the linear coupling 
approximation is not valid even in systems where the coupling is weak, and 
that higher order couplings strongly influence the calculated barrier 
distributions. 

\begin{flushleft}
{\bf 2. Coupled channels equations in all orders}
\end{flushleft}

Consider the problem where the relative motion between  
colliding nuclei couples to a vibrational mode of excitation of the target 
nucleus. For simplicity excitations of the 
projectile are not considered in this section. It is straightforward 
to extend the formulae to the case 
where many different vibrational modes are present and where projectile 
excitations also occur. For heavy ion fusion reactions, 
to a good approximation one can replace the angular momentum of the relative 
motion in each channel by the total angular momentum $J$\cite{HTBBe95}. 
This approximation, often referred to as no-Coriolis approximation, will be 
used throughout this paper. The coupled channels equations then read 
\begin{equation}
\left[-\frac{\hbar^2}{2\mu}\frac{d^2}{dr^2}
+\frac{J(J+1)\hbar^2}{2\mu r^2}
+\frac{Z_PZ_Te^2}{r}+n\hbar\omega
-E_{cm}\right]\psi_n(r)+\sum_mV_{nm}(r)\psi_m(r)=0  ,
\end{equation}
where $r$ is the radial component of the coordinate of the 
relative motion, $\mu$ the reduced mass, 
$E_{cm}$ the 
bombarding energy in the center of mass frame and 
$\hbar\omega$ is the excitation energy of 
the vibrational phonon. 
$V_{nm}$ are the coupling matrix elements, which in the collective model 
consist of Coulomb and nuclear components. 

For the Coulomb coupling, it has been shown that inclusion of up to 
the first order term is sufficient\cite{HTDHL97}. 
The matrix elements of the Coulomb coupling in Eq. (1) thus read 
\begin{equation}
V^{(C)}_{nm}(r)=\frac{3}{2\lambda+1}Z_PZ_Te^2
\frac{R_T^{\lambda}}{r^{\lambda+1}}
\sqrt{\frac{2\lambda+1}{4\pi}}\alpha_0 (\sqrt{n}\delta_{n,m+1}
+\sqrt{n+1}\delta_{n,m-1}),
\end{equation}
where $\alpha_0$ is the amplitude of the zero point motion. 
It is related to the deformation parameter $\beta_{\lambda}$ 
by $\alpha_0=\beta_{\lambda}/\sqrt{2\lambda+1}$\cite{BM75}. 

In the collective model, the nuclear interaction is assumed to be 
a function of the separation distance between the vibrating surfaces 
of the colliding nuclei. 
It is conventionally taken as 
\begin{equation}
V^{(N)}(r,\alpha_{\lambda0})
=-\frac{V_0}{1+\exp[(r-R_P-R_T-\sqrt{\frac{2\lambda+1}{4\pi}}
R_T\alpha_{\lambda0})/a]}, 
\end{equation}
where the surface 
coordinates $\alpha_{\lambda\mu}$ are 
related to the phonon creation and annihilation operators by 
%
$\alpha_{\lambda\mu}=\alpha_0(a^{\dagger}_{\lambda\mu}
+(-)^{\mu}a_{\lambda-\mu})$. 
%
Volume conservation introduces a small term which is non-linear with respect 
to the deformation parameter $\alpha_{\lambda0}$ in the denominator of the 
above Eq.(3). This is ignored for simplicity in the present study. 

The conventional nuclear coupling form factor in the linear coupling 
approximation is obtained by expanding Eq.(3) with respect to 
$\alpha_{\lambda0}$ and keeping only the linear term. 
It is given by 
\begin{equation}
V^{(N)}_{nm}(r)=V_N(r)\delta_{n,m}-
\sqrt{\frac{2\lambda+1}{4\pi}}\alpha_0 
\frac{dV_N(r)}{dr} (\sqrt{n}\delta_{n,m+1}
+\sqrt{n+1}\delta_{n,m-1}),
\end{equation}
where $V_N(r)=-V_0/[1+\exp((r-R_p-R_T)/a)]$ is the nuclear potential 
in the entrance channel. 

In order to take into account the effects of the couplings to all order, 
we evaluate the nuclear coupling matrix elements without introducing 
the expansion. 
Denoting the eigenvalue of $\alpha_{\lambda0}$ by $x$, 
the matrix elements of the nuclear coupling then read 
\begin{equation}
V_{nm}^{(N)}(r)
=\int^{\infty}_{-\infty}dxu^*_n(x)u_m(x)
\frac{-V_0}{1+\exp[(r-R_P-R_T-\sqrt{\frac{2\lambda+1}{4\pi}}
R_T x)/a]}~~.
\end{equation}
Here $u_n(x)$ is the eigen function of the $n$ -th excited state of the 
harmonic oscillator and is given by 
\begin{equation}
u_n(x)=\frac{1}{2^nn!}\frac{1}{\sqrt[4]{2\pi\alpha_0^2}}
H_n\left(\frac{x}{\sqrt{2}\alpha_0}\right)e^{-x^2/4\alpha_0^2}, 
\end{equation}
where $H_n(x)$ is the Hermite polinomial with rank $n$.

\begin{flushleft}
{\bf 3. Validity of the linear coupling approximation}
\end{flushleft}

\begin{flushleft}
{\it 3.1. Nearly Symmetric Systems}
\end{flushleft}

We now present the results of our calculations of fusion cross sections, 
average angular momenta of the compound nucleus, and fusion barrier 
distributions. 
We first discuss heavy nearly symmetric systems. 
We analyse in particular $^{64}$Ni + $^{96}$Zr 
reactions which are typical examples where 
the conventional coupled channels calculations with the linear 
coupling approximation fail to reproduce the 
fusion cross sections and average angular momentum data\cite{SCA92}. 
Our aim is to investigate whether 
the failure is due to the linear coupling approximation by 
performing linear, quadratic and full coupling calculations. 

We take into account the couplings up to two phonon states 
of the quadrupole surface vibrations of $^{64}$Ni, 
and of the octupole vibration of 
$^{96}$Zr. We also take their mutual excitations into account. 
We ignore the effects of transfer reactions, because  
it has been reported in Ref.\cite{SCA92} that they have only 
small effects on the fusion cross sections and the average angular 
momenta in this reaction. 
The excitation function of the fusion cross section for this reaction 
obtained by numerically solving 
the coupled channels equations is compared with the experimental data 
in Fig. 1 (upper panel). The experimental data, 
taken from Ref. \cite{SCA92}, consist only 
of the evaporation residue cross sections, 
and do not include fission following fusion. 
The dotted line is the results in the one dimensional 
potential model, {\it i.e.} without the effects of channel coupling. 
The dot-dashed line is the results of the coupled 
channels calculations when the linear coupling approximation is used. 
They considerably underestimate the fusion cross sections 
at sub-barrier energies. 
The situation is slightly improved when the 
quadratic coupling approximation\cite{EL87} is used, {\it i.e.} when 
the nuclear coupling potential up to the second order of the deformation 
parameter, is included (dashed line).  
However, there still remain considerable discrepancies between the 
experimental data and the results of the coupled channels calculations. 
When we include  couplings to all order, we get the solid line, which 
agree very well with the experimental data. Dramatic effects 
of the higher order couplings on fusion cross sections are 
observed, especially at low energies. 

The lower panel in Fig. 1 compares the 
results of our calculations of the average angular momentum 
of the compound nucleus with the experimental data 
as a function of the bombarding energy.
We again observe that the experimental data are much better reproduced 
by taking the effects of couplings to all orders into account. 
We thus conclude that coupling to all 
orders are essential to simultaneously reproduce the fusion cross 
sections and the average angular momentum data 
for heavy (nearly) symmetric systems. This is in agreement with the 
calculations required to fit the barrier distribution for 
$^{58}$Ni + $^{60}$Ni reaction\cite{SACN95}. 

\begin{flushleft}
{\it 3.2. Very Asymmetric Systems}
\end{flushleft}

We next consider the effects of higher order couplings for 
very asymmetric systems where the product of the charges $Z_PZ_T$  
is relatively small. 
For such systems, the coupled channels calculations in the linear 
coupling approximation have achieved reasonable success 
in reproducing fusion excitation functions. 
However, no study has been performed to see whether 
the effects of higher order couplings on the angular 
momentum distribution of the compound nucleus and on the barrier 
distributions are small. In this subsection we re-analyse 
the experimental data 
for the $^{16}$O + $^{144}$Sm reaction, 
for which the effects of couplings to 
phonon states on fusion barrier distributions were shown experimentally 
for the first time\cite{LDH95}.

The fusion barrier distribution for this system is 
intimately related to the octupole vibration of $^{144}$Sm, and that 
the quadrupole vibration plays only a minor role\cite{LDH95}. 
Accordingly, we ignore the effects of the 
couplings to the quadrupole phonon states of $^{144}$Sm and include only the 
single octupole phonon state at 1.81 MeV. 
For simplicity in the calculations we ignore excitation of the projectile. 
These effects will be discussed in Sec. 4. 

The upper panel of Fig. 2 shows the experimental fusion excitation function 
taken from Ref. \cite{LDH95}, and the theoretical calculations.  
The meaning of each line 
is the same as in Fig.1. We observe that the 
agreement of the theory and experiment 
appears to be improved only slightly by the inclusion of coupling to all 
orders. 
The barrier distribution, on the other hand, 
reveals significant changes due to 
the higher order couplings (see the lower panel of Fig. 2). 
Comparing the results of the linear coupling approximation (the dot-dashed 
line) with those of the all order coupling (the solid line), one observes 
that the higher order couplings  
transfer some strength from the lower barrier 
to the higher barrier, 
and at the same time lower the peak position of both barriers. 

Figure 3 shows the coupling matrix element between the ground state and the 
one phonon state $V_{01}(r)$ as a function of the inter nuclear separation 
distance $r$. The dotted line is the coupling matrix element in the 
linear coupling approximation, while the solid line includes the coupling 
to all orders. 
One can see that the linear coupling approximation underestimates 
the coupling 
strength at the barrier position of the uncoupled barrier 
around $r$= 10.8 fm. 
On the other hand, it overestimates the coupling strength in the inner region 
around $r$=8.5 fm. As we will see in the next section, the latter fact is 
important in discussing the effects of the projectile excitation. 

The effects of higher order couplings become more significant 
when there exist more than two channels. 
In order to demonstrate this, we show in fig. 4 the barrier distributions for 
the $^{16}$O + $^{144}$Sm reaction, where the double octupole phonon 
excitations are allowed in the harmonic limit. 
One observes dramatic effects of higher order couplings 
on the fusion barrier distribution. 
It is thus clear 
that high precision measurements should be analysed using all 
order couplings even when coupling is weak as a result of the small charge 
product. 
The detailed studies on the effects of double phonon couplings including 
anharmonic effects in this reaction is reported separately \cite{HTK97,THK97}. 

\begin{flushleft}
{\bf 4. Role of projectile excitation in fusion}
\end{flushleft}

Let us now discuss the role of the octupole vibrational 
excitation of $^{16}$O in the $^{16}$O + $^{144}$Sm fusion 
reactions\cite{HTDHL97b}. 
Contradictory conclusions have been reported regarding the role of 
projectile excitation in the fusion reactions between $^{16}$O 
and samarium isotopes. 
Calculations of the fusion cross-section for the $^{16}$O + 
$^{154}$Sm reactions in ref. \cite{GCW94} indicated the 
importance of the excitation of $^{16}$O. In marked contrast, 
no specific features appear 
in the measured barrier distribution for the $^{16}$O + $^{144}$Sm 
reactions which can be associated with the excitation 
of $^{16}$O; rather, it was argued in Ref. \cite{LDH95} that a good 
theoretical representation of the experimental fusion 
barrier distribution is destroyed when 
the projectile excitation is included. 

Both of these conclusions are based on the comparison of the experimental 
data with the results of simplified coupled-channels calculations, where 
the linear coupling approximation is used to describe the vibrational 
excitation of the projectile. 
In the previous section, we have shown that the linear coupling 
approximation is not valid even in systems with weak coupling, and 
that higher order couplings strongly influence the fusion barrier 
distribution. 
The coupling to the octupole vibrational state of $^{16}$O is 
strong because of the large deformation parameter indicated by the 
strong E3 transition. It is therefore very likely that 
the fusion barrier distribution calculated 
in the simplified coupled--channel codes 
using the linear coupling approximation  
does not represent the true fusion barrier distribution. 

In order to test the validity of the simplified coupled-channels 
calculations, here we first calculate the excitation function of 
the fusion cross section and the fusion barrier distribution 
in the linear coupling approximation. 
The results are shown in Fig.~5. 
The dotted line is the results when $^{16}$O is treated 
to be inert. This calculation 
well reproduces the features of the experimental barrier distribution. 
The results of calculations including  the excitation of the lowest-lying 
octuple state of $^{16}$O are shown by the solid line. Though 
the experimental barrier distribution around 
the lower energy peak at ($\sim$ 60 MeV) is reproduced, 
significant strength is 
missing around the higher energy peak near 65 MeV. 
A similar discrepancy between theory and experimental data 
was encountered in Ref.~\cite{LDH95}, where calculations 
were performed using a modified version of the CCFUS 
code (the long--dashed line). 
Clearly both calculations which treat the 
octupole excitation of $^{16}$O in the linear coupling approximation  
fail to reproduce the experimental barrier distribution.  

The results of coupled-channels calculations, where
the couplings to the octupole vibrations of both $^{16}$O and $^{144}$Sm
are treated to all orders, are shown in figure~6.
It is remarkable that these calculations
re-establish the double-peaked structure seen in the experimental data,
which was missing in the linear coupling calculations.
Indeed,
the barrier distribution obtained by
including the coupling to the octupole vibration of $^{16}$O
to all order looks very similar to that obtained by totally ignoring it,
apart from a shift in energy.  A shift of 2 MeV (dashed line) of the former
is required for the two calculated distributions to coincide.
This shift
is consistent with the general conclusion that the main effect of
the coupling to an inelastic channel whose excitation energy is larger
than the curvature of the bare fusion barrier, i.e. an adiabatic
coupling, is to introduce a static potential shift\cite{THAB94},
and hence, the shape of the barrier distribution does not change
unless the coupling is very strong and the coupling form factor
itself has a strong radial dependence.
The effects of these excitations can then be included in the `bare' potential
in the coupled-channels calculations.
Thus for
$^{16}$O~+~$^{144}$Sm, where potential parameters for the
calculations are obtained from
a fit to the high energy data, the effects of octupole vibration of
$^{16}$O are already included. The explicit inclusion of the coupling to
octupole vibration then leads to double counting which manifests itself
by introducing an additional shift in the barrier (or barrier distribution)
as observed earlier.

\begin{flushleft}
{\bf 5. Summary}
\end{flushleft}

We have shown that  higher order couplings 
to nuclear surface vibrations play an important role 
in heavy ion fusion reactions. 
The inclusion of the coupling to all orders 
in coupled-channels calculations was shown to be 
crucial to reproduce the experimental 
fusion cross sections and the average angular momenta for 
heavy symmetric systems. 
We performed coupled-channels calculations also for the $^{16}$O + $^{144}$Sm 
reactions as an example of very asymmetric systems where the coupling 
is weaker. 
It was found that higher order couplings to the 
vibrational states of the target nucleus result 
in a non-negligible enhancement of the  fusion cross sections and a 
significant modification of barrier distributions 
even in very asymmetric systems. 
Our studies warn that spurious conclusions could be reached regarding 
the nature of couplings if high quality experimental data 
are compared with simplified calculations in the first 
order approximation. 
We also studied the role of projectle excitation in the 
$^{16}$O + $^{144}$Sm fusion reactions.
Using the calculations with full order coupling 
we have shown that the major effect of the excitation of the 
octupole state at 6.1 MeV of $^{16}$O  
is to renormalize the static 
potential barrier without significantly 
modifying the shape of the barrier distribution. 

\begin{flushleft}
{\bf Acknowledgments} 
\end{flushleft}

The authors thank S. Kuyucak for useful discussions. 
The work of K.H. was supported by the Japan Society for the Promotion 
of Science for Young Scientists.
This work was supported by the Grant-in-Aid for General
Scientific Research,
Contract No.06640368 and No.08640380, and the Grant-in-Aid for Scientific
Research on Priority Areas, Contract No.05243102 and 08240204  
from the Japanese Ministry of Education, Science and Culture, 
and a bilateral program of JSPS between Japan and Australia.


\begin{center}
{\bf Figure Captions}
\end{center}

\noindent
{\bf Fig.1:}Excitation function of the fusion cross section (upper 
panel) and the average angular momentum of the compound nucleus 
(lower panel) for the 
$^{64}$Ni + $^{92}$Zr reactions. 

\noindent
{\bf Fig.2:}Excitation function of the fusion cross section (upper 
panel) and the fusion barrier distribution (lower panel) for 
the $^{16}$O + $^{144}$Sm reactions. In the coupled channels calculations, 
the projectile is assumed to be inert, while the single octupole 
phonon state of the target nucleus is taken into account. 

\noindent
{\bf Fig.3:}The coupling matrix element between the ground state and the 
one phonon state for the $^{16}$O + $^{144}$Sm reaction as 
a function of the separation distance between the projectile and 
target.. 

\noindent
{\bf Fig.4:}Same as the lower panel of fig. 2, 
but for the case where the double octupole phonon 
excitations of $^{144}$Sm are included in the harmonic limit. 

\noindent
{\bf Fig.5:} Effects of the projectile excitation on 
the $^{16}$O + $^{144}$Sm fusion reactions. 
The linear coupling approximation is used 
in the coupled-channels calculations. 
In all calculations, the effects of the octupole vibration of 
$^{144}$Sm are taken into account.

\noindent
{\bf Fig.6:} Same as Fig.5, but for the case where the 
coupled-channels calculations have been performed to 
all order coupling.


\newpage

\begin{thebibliography}{99} 

\bibitem{LDH95}J.R. Leigh, M. Dasgupta, D.J. Hinde, J.C. Mein, 
C.R. Morton, R.C. Lemmon, J.P. Lestone, J.O. Newton, H. Timmers, 
J.X. Wei, and N. Rowley, 
Phys. Rev. C{\bf 52}, 3151(1995), and references there in.

\bibitem{RSS91}N. Rowley, G.R. Satchler, and P.H. Stelson, 
Phys. Lett. {\bf B254}, 25(1991). 

\bibitem{CBD95}A. Charlop, J. Bierman, Z. Drebi, A. Garc\'ia, S. Gil, 
D. Prindle, A. Sonzogni, R. Vandenbosch, and D. Ye, Phys. Rev. C{\bf 51}, 
628(1995). 

\bibitem{HTDHL97}K. Hagino, 
N. Takigawa, M. Dasgupta, D.J. Hinde, and
J.R. Leigh, Phys. Rev. C{\bf 55}, 276(1997). 

\bibitem{BBK94}A.B. Balantekin, J.R. Bennett, and S. Kuyucak,
Phys. Rev. C{\bf 48}, 1269(1993); {\bf49}, 1079(1994).

\bibitem{HTBBe95}K. Hagino, N. Takigawa, A.B. Balantekin, and 
J.R. Bennett, Phys. Rev. C{\bf 52}, 286(1995). 

\bibitem{BM75}A. Bohr and B.R. Mottelson, {\it Nuclear Structure}, vol.2, 
(Benjamin, Reading, 1975).

\bibitem{SCA92}A.M. Stefanini, L. Corradi, D. Ackermann, A. Facco, 
F. Gramegna, H. Moreno, L. Mueller, D.R. Napoli, G.F. Prete, 
P. Spolaore, S. Beghini, D. Fabris, G. Montagnoli, 
G. Nebbia, J.A. Ruiz, G.F. Segato, C. Signorini, and G. Viesti, 
Nucl. Phys. {\bf A548}, 453(1992). 

\bibitem{EL87}H. Esbensen and S. Landowne, Phys. Rev. C{\bf 35}, 
2090(1987). 

\bibitem{SACN95}A.M. Stefanini, D. Ackermann, L. Corradi, D.R. Napoli, 
C. Petrache, P. Spolaore, P. Bednarczyk, H.Q. Zhang, S. Beghini, 
G. Montagnoli, L. Mueller, F. Scarlassara, G.F. Segato, F. Sorame, 
and N. Rowley, 
Phys. Rev. Lett. {\bf 74}, 864(1995). 

\bibitem{HTK97}K. Hagino, N. Takigawa, and S. Kuyucak, (to be published). 

\bibitem{THK97}N. Takigawa, K. Hagino and S. Kuyucak, contribution 
to this conference. 

\bibitem{HTDHL97b}K. Hagino, 
N. Takigawa, M. Dasgupta, D.J. Hinde, and
J.R. Leigh, (to be published). 

\bibitem{GCW94}P.R.S. Gomes, 
I.C. Charret, R. Wanis, G.M. Sigaud, 
V.R. Vanin, R. Liguori Neto, D. Abriola, O.A. Capurro, 
D.E. DiGregorio, M. di Tada, G. Duchene, M. Elgue, A. Etchegoyen, 
J.O. Fern\'andez Niello, A.M.J. Ferrero, S. Gil, A.O. Macchiavelli, 
A.J. Pacheco, and J.E. Testoni, 
Phys. Rev. C{\bf 49}, 245(1994).

\bibitem{THAB94}N. Takigawa, K. Hagino, M. Abe and A.B. Balantekin,
Phys. Rev. C{\bf 49}, 2630(1994); 
N. Takigawa, K. Hagino, and M. Abe, Phys. Rev. C{\bf 51}, 
187(1995).

\end{thebibliography}
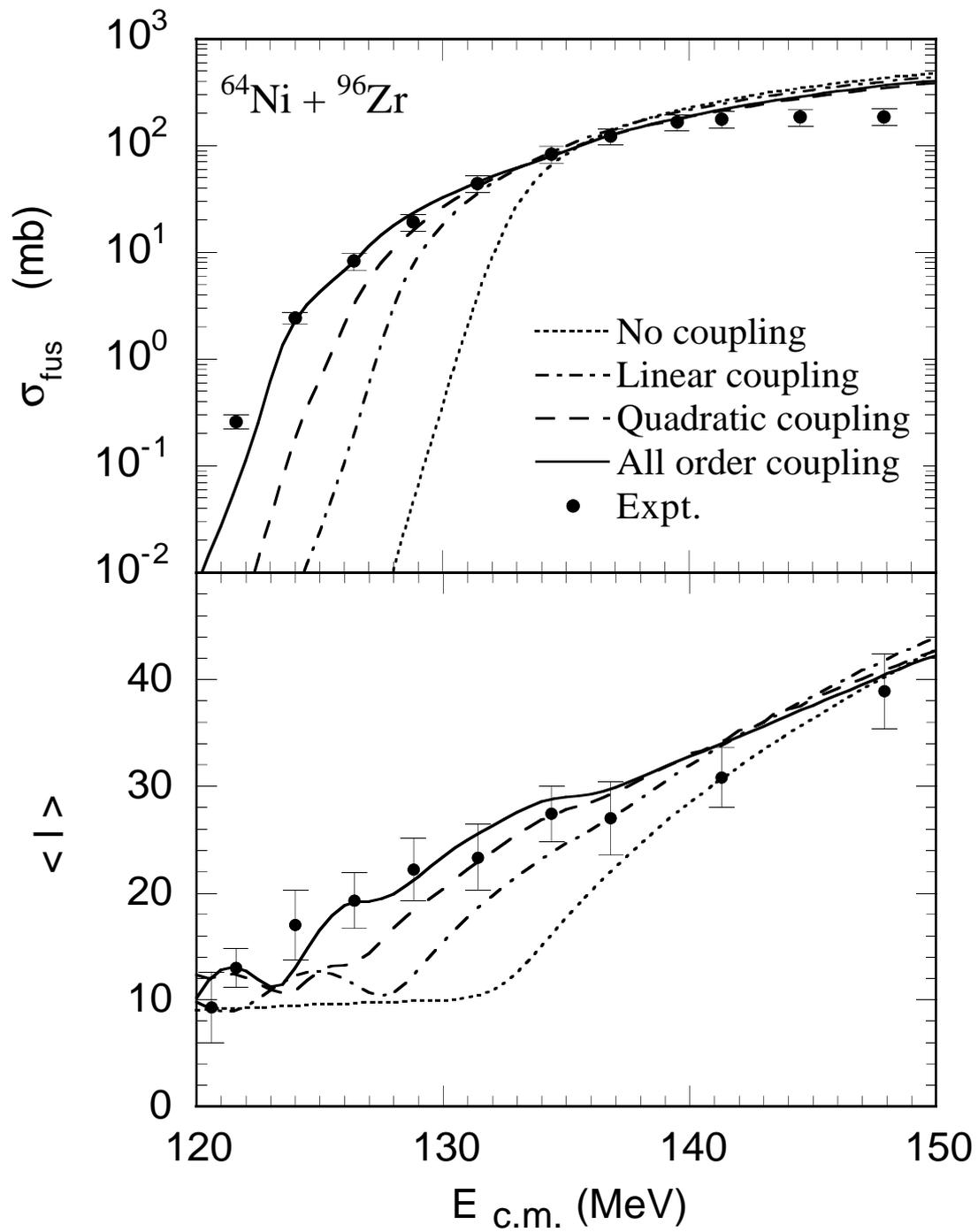
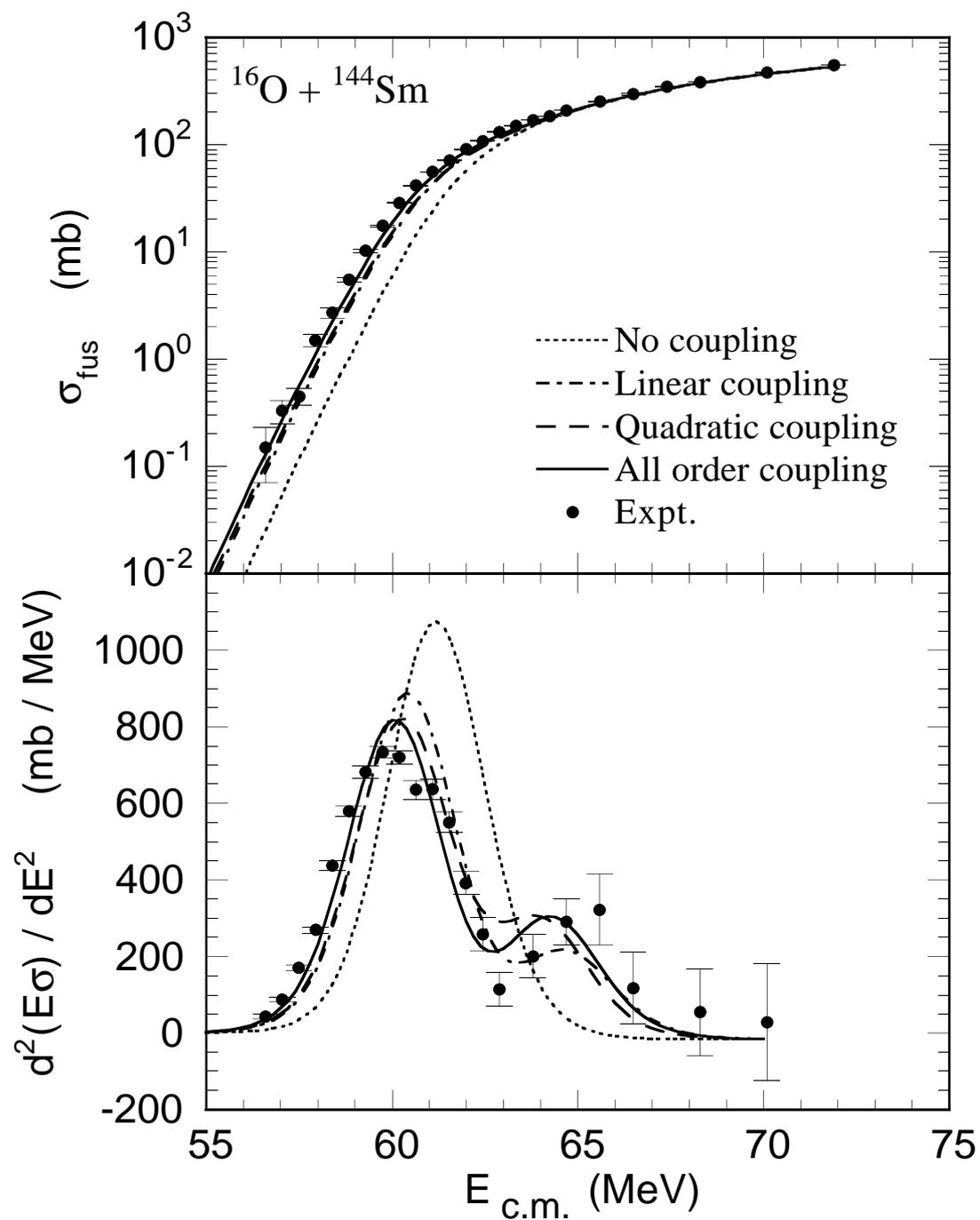
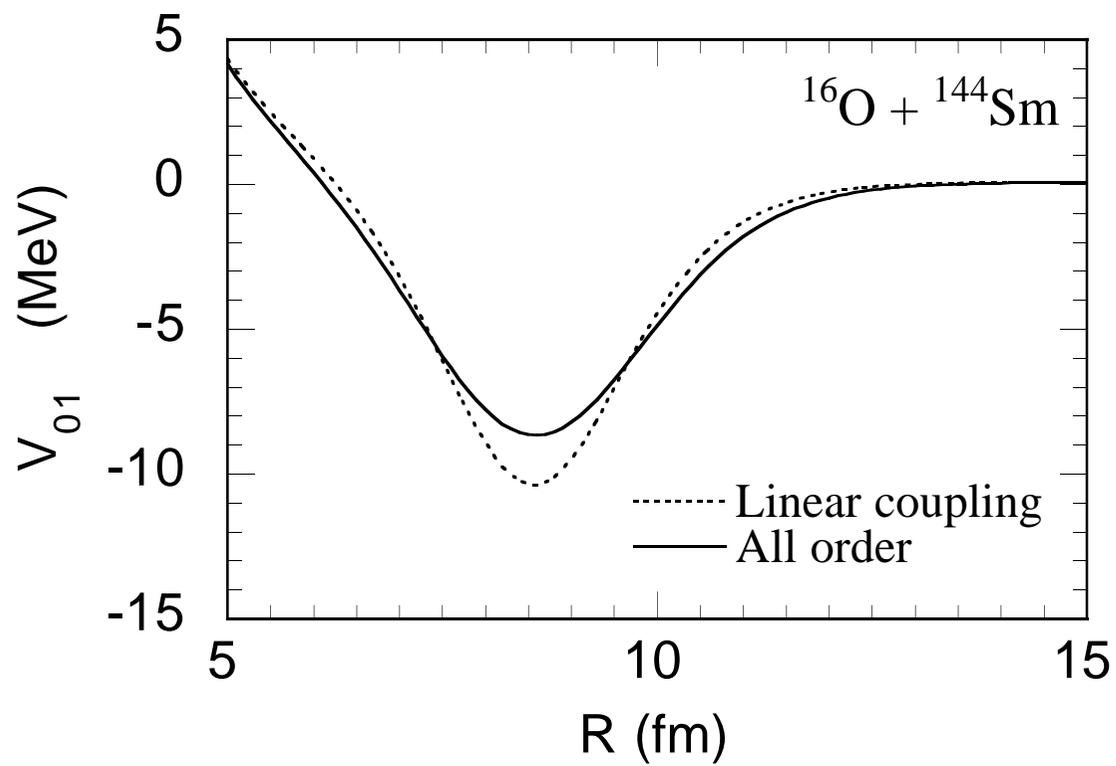
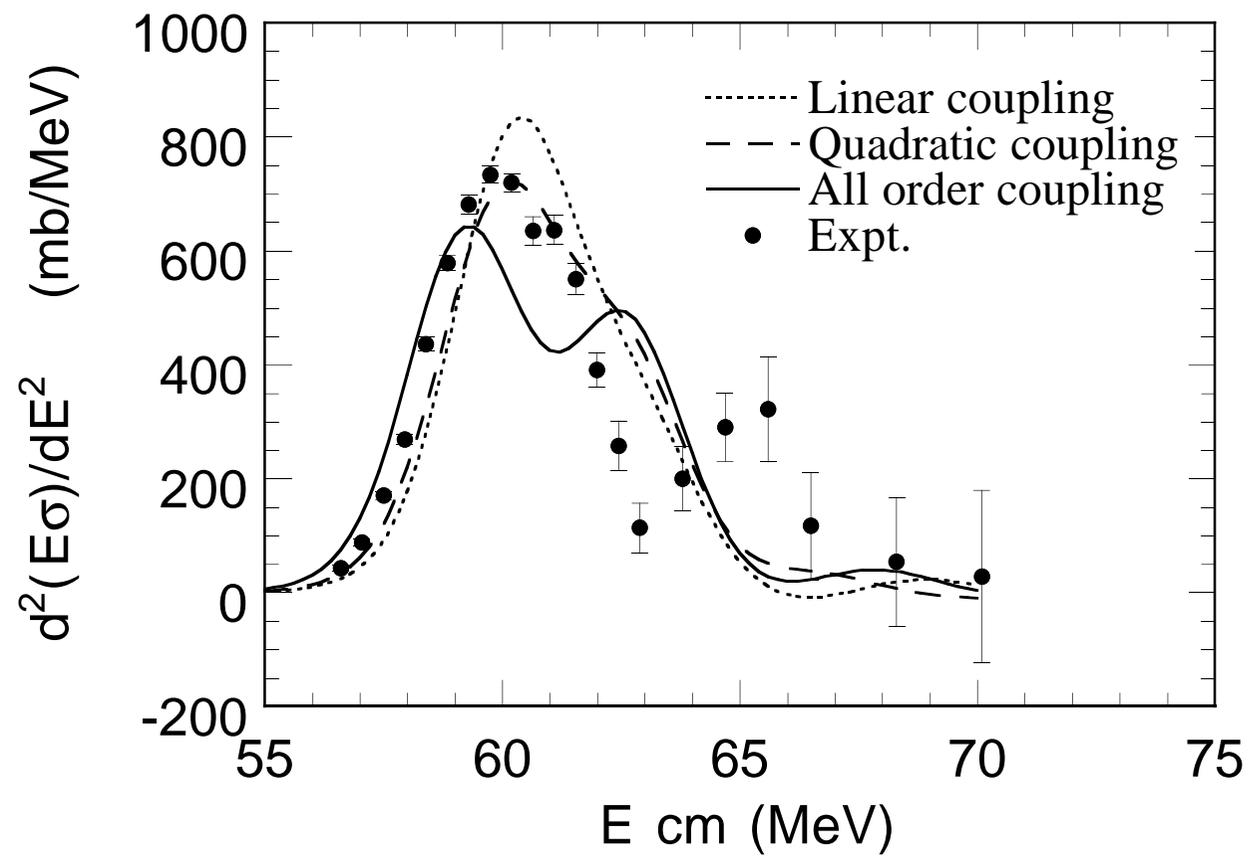
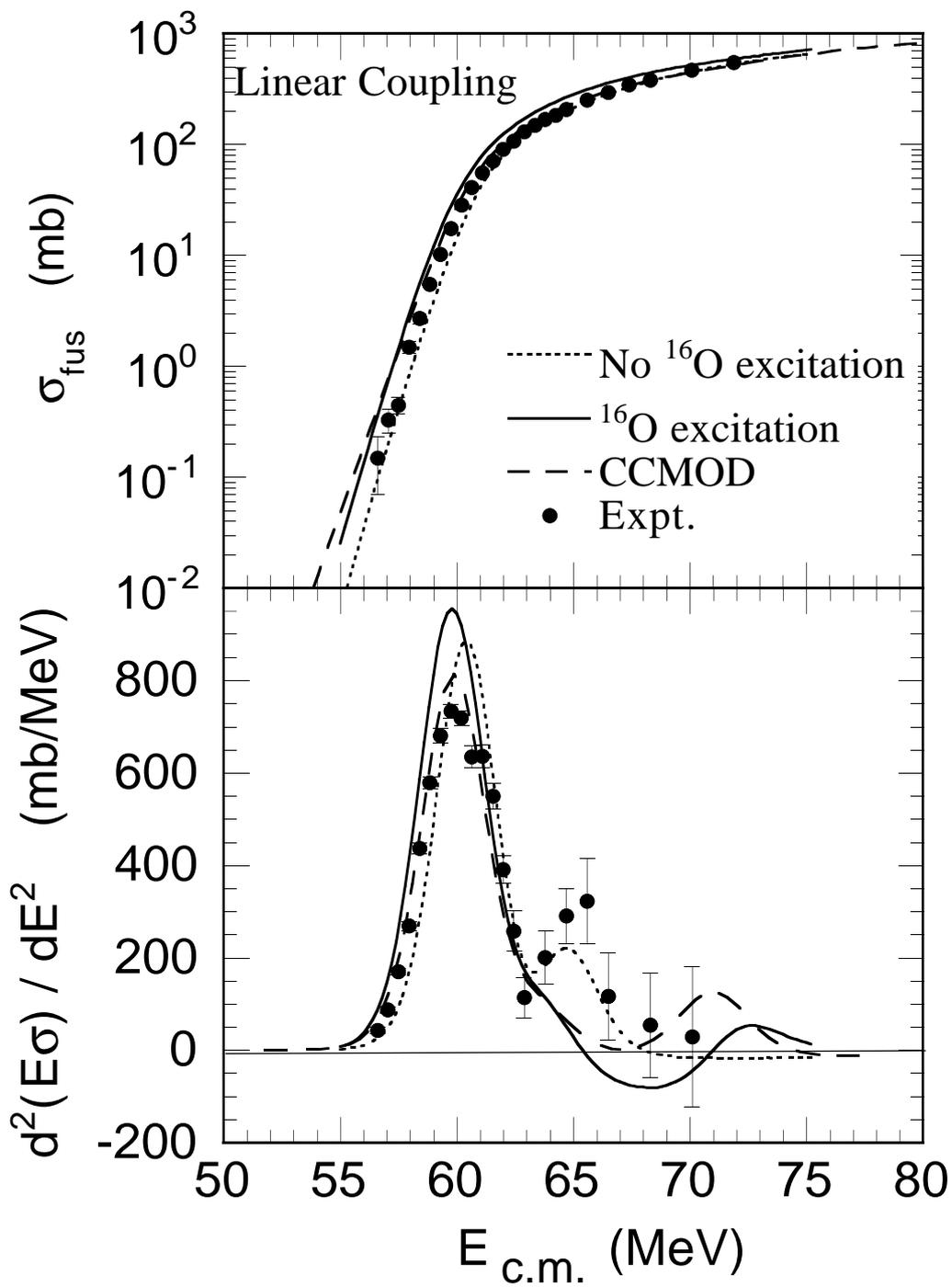
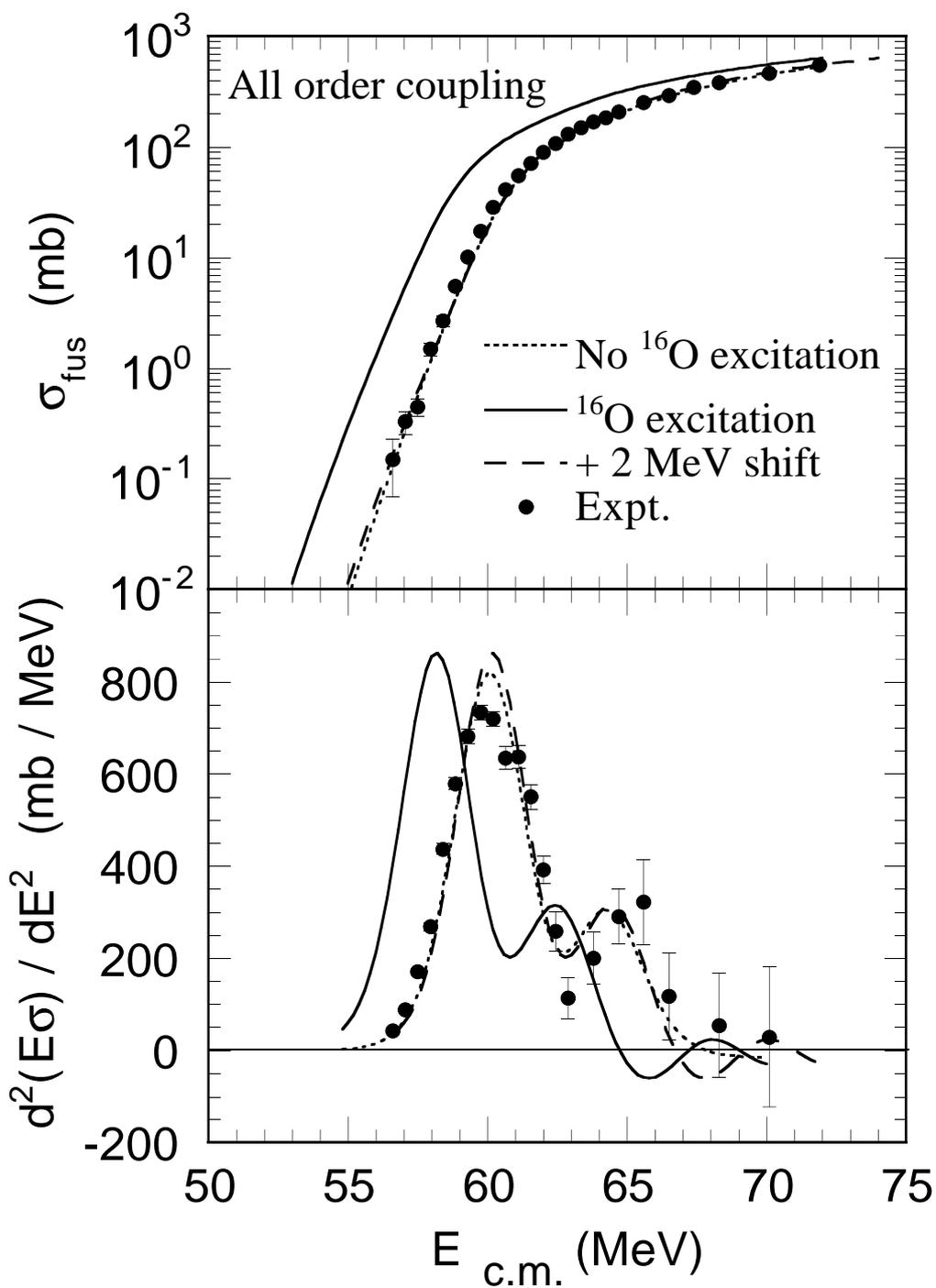
\end{document}